\documentclass[bm,aps,showpacs,amsmath,preprint,amssymb,12pt]{revtex4}
\usepackage{amsmath,amssymb,bm}

\usepackage{graphicx}

\begin{document}
{~}
\vspace{3cm}

\title{Charged Black Holes in a Rotating Gross-Perry-Sorkin 
Monopole Background 

}
\author{Shinya Tomizawa$^{1}$ and Akihiro Ishibashi$^{1,2}$}
\affiliation{
${}^{1}$Cosmophysics Group, Institute of Particle and Nuclear Studies, 
KEK, Tsukuba, Ibaraki, 305-0801, Japan \\ 
and \\ 
${}^{2}$Perimeter Institute for Theoretical Physics, Waterloo, 
ON N2L 2Y5, Canada 
}
\begin{abstract}
We present a new class of stationary charged black hole solutions to 
five-dimensional Einstein-Maxwell-Chern-Simons theories. We construct 
the solutions by utilizing so called the squashing transformation. 
At infinity, our solutions behave as a four-dimensional flat spacetime 
plus a `circle' and hence describe a Kaluza-Klein black hole.   
More precisely, our solutions can be viewed as a charged rotating black 
hole in a rotating Gross-Perry-Sorkin monopole background with the black hole 
rotation induced from the background rotation. 
\end{abstract}

\date{\today}

\preprint{KEK-TH-1261}

\maketitle

\section{Introduction}

In recent years, higher dimensional black holes have become one of 
the major subjects in fundamental physics~\cite{MP,ER,MI,PS,F,IM_BD,EF,BMPV,SUSYRing,SUSYRing2,C_ring,C_ring2,CLP,Cai,Helfgott,Galloway,Morisawa,MTY,HIW}.  
Although recent 
ideas of braneworld and TeV gravity~\cite{ADD,RS1,RS2} 
have opened up the possibility of `large' extra-dimensions, still important 
and widely believed especially in the context of string theories 
is the spacetime picture that our macroscopically large four-dimensional 
spacetime is realized from a higher dimensional 
spacetime by some mechanism of compactifying (and stabilizing) 
extra-dimensions within the size of the fundamental scale of gravity. 
In this context, it is of great interest to consider 
{\sl Kaluza-Klein black holes}, which are essentially higher dimensional 
near the event horizon but look like four-dimensional with compactified 
small extra-dimensions, at large distances.

\medskip 
The purpose of this paper is to provide a new class of Kaluza-Klein black 
hole solutions, which are constructed by applying so called 
the {\sl squashing transformation} (described below) 
to a known {\sl starting-point} 
solution in five-dimensional theories. 
Specifically, we consider the Reissner-Nordst\"om-G\"odel black hole 
solution of Herdeiro~\cite{Herdeiro} 
in five-dimensional Einstein-Maxwell-Chern-Simons theories,  
as our starting-point geometry. We ``squash'' it, thereby obtaining 
a new metric specified by 4 parameters.  
We then choose the parameters so that the resultant 
spacetime possesses two Killing horizons. 
As a result, our new solutions have the black hole event horizon 
whose cross-section geometry is of squashed $\rm S^3$, as well as an inner 
horizon, and are asymptotically locally flat with the structure of 
a twisted $\rm S^1$ bundle over a four-dimensional Minkowski spacetime.  
It is remarkable to note that although our starting-point 
geometry~\cite{Herdeiro} admits closed timelike curves as a consequence of 
G\"odel type rotation, our resultant new Kaluza-Klein solution itself 
exhibits no causal pathology outside the event horizon, thanks to 
the squashing transformation and appropriate choice of the parameters.

\medskip 
Our Kaluza-Klein black holes obtained in the way described above 
can also be viewed as a charged rotating black hole in the rotating 
Gross-Perry-Sorkin (GPS) monopole background~\cite{GP,TIMN}, which itself is 
obtained via the squashing transformation from the five-dimensional 
G\"odel universe in Einstein-Maxwell-Chern-Simons theories. 
Therefore, in order to understand basic properties of our Kaluza-Klein 
solutions, it would be illuminating to discuss first the geometry and 
basic features of the rotating GPS background solutions. 
We shall do so in the next section. But before proceeding subsequent sections 
and presenting our solutions, we would like to review briefly some related 
work.  

\medskip 
The first example of such Kaluza-Klein black holes was given as a vacuum 
solution to the five-dimensional Einstein equation by Dobiasch and 
Maison~\cite{DM}. It was then shown by Gibbons and Wiltshire~\cite{GW} 
that the solution of Dobiasch and Maison asymptotes to a twisted $\rm S^1$ 
bundle over the four-dimensional spacetime with a single rotation with respect 
to the extra fifth-dimension. (See e.g., Rasheed~\cite{Rasheed} for further 
generalization of the Dobiasch and Maison solution~\cite{DM}.)  

\medskip 
A recent key work along this line is that of Ishihara and Matsuno~\cite{IM}, 
who have found static charged Kaluza-Klein black hole solutions in the 
five-dimensional Einstein-Maxwell theory by using, for the first time, 
the squashing technique. 
Namely, they view the base manifold $\rm S^3$ of a five-dimensional 
Reissner-Nordst\"om black hole as a fiber bundle over $\rm S^2$ with 
fiber $\rm S^1$ and then consider a deformation that changes the ratio of 
the radius of $\rm S^2$ to that of $\rm S^1$ fiber. 
Subsequently the Ishihara-Matsuno solution was 
generalized to many different cases. For example, a static multi Kaluza-Klein 
black hole solution~\cite{IKMT} was constructed immediately from 
the Ishihara-Matsuno solution.  
The squashing transformation of Ishihara-Matsuno~\cite{IM}, which we will also 
employ in this paper, has been then recognized as a type of powerful solution 
generating technique. In fact, it was demonstrated by Wang~\cite{Wang} 
that the five-dimensional Kaluza-Klein black hole of Dobiasch and Maison can 
be reproduced by squashing a five-dimensional Myers-Perry black hole 
with two equal angular momenta.

\medskip 
The application of the squashing transformation to non-asymptotically 
flat Kerr-G\"odel black hole solutions~\cite{Gimon-Hashimoto} was 
considered in \cite{TIMN}.    
In \cite{NIMT}, the squashing transformation was applied 
to the Cvetic {\it et al}'s charged rotating black hole 
solution~\cite{CLP} with two equal angular momenta.  
These solutions \cite{NIMT,TIMN} correspond to 
a generalization of the Ishihara-Matsuno solution to the rotating 
black holes in Einstein-Maxwell-Chern-Simons theories. 
A similar type of Kaluza-Klein black holes was considered in the context of 
supersymmetric theories by Gaiotto {\it et al}~\cite{Gaiotto} 
and Elvang {\it et al}~\cite{EEMH}.  
Kaluza-Klein black holes which asymptote to the direct product 
of the four-dimensional Minkowski space-time and $\rm S^1$ were discussed 
by ~\cite{Myers,TIM,IMT}.   
Further generalizations of Kaluza-Klein solutions of this type and 
their basic properties have been studied by many 
authors~\cite{BKW,HO,KRT,IKMT2,IKT,MIKT,TIKM,T,IIKMMT,YIKMT}.

\medskip 
The rest of this paper is organized as follows. 
In the next section, as mentioned above, we will review the five 
dimensional G\"odel universe and the rotating GPS monopole based 
on the results of the paper~\cite{TIMN}. 
Then, in section \ref{sec:solution}, we present our new stationary charged 
Kaluza-Klein black hole solutions in the rotating GPS monopole background. 
In section \ref{sec:case}, we examine our solutions under several special 
choices of the parameters and study their basic properties. 
Section \ref{sec:summary} is devoted to summary and discussions. 
In Appendix we show that our background, rotating GPS monopole geometry 
is a supersymmetric solution.

\section{Rotating Gross-Perry-Sorkin monopole}\label{sec:GPS}
We shall briefly review the results of~\cite{TIMN} concerning the 
{\sl rotating Gross-Perry-Sorkin (GPS) monopole}, which is one of the simplest 
solutions in five-dimensional Einstein-Maxwell-Chern-Simons theories.  
We begin with the five-dimensional G\"odel universe since the rotating GPS 
monopole solutions are obtained by applying the squashing transformation 
to the five-dimensional supersymmetric G\"odel universe.  
We then discuss some basic properties of the 
rotating GPS monopole, in particular, its asymptotic structure 
and the existence of an ergoregion. 

\subsection{Five-dimensional G\"odel universe} 
Consider the five-dimensional Einstein-Maxwell theory with a Chern-Simons  
term, whose action is given by  
\begin{eqnarray}
 S = \frac{1}{16 \pi G_5} 
    \int d^5 x \left[ 
                    \sqrt{-g} \left( R - F_{\mu \nu } F^{\mu \nu } \right) 
                    + \frac{2}{3 \sqrt 3} 
                      \epsilon ^{\mu \nu \rho \sigma \lambda } 
                      A_\mu F_{\nu \rho } F_{\sigma \lambda } 
               \right] \,, 
\label{action}
\end{eqnarray}
where $R$ is the five dimensional scalar curvature, 
$F = dA$ is the two-form of the five-dimensional gauge field 
associated with the gauge potential one-form $A$, and $G_5$ is 
the five-dimensional Newton constant. Varying the action (\ref{action}), 
we derive the Einstein equation 
\begin{eqnarray}
 R_{\mu \nu } -\frac{1}{2} R g_{\mu \nu } 
 = 2 \left( F_{\mu \lambda } F_\nu^{ ~ \lambda } 
  - \frac{1}{4} g_{\mu \nu } F_{\rho \sigma } F^{\rho \sigma } \right) \,, 
 \label{Eineq}
\end{eqnarray}
and the Maxwell equation 
\begin{eqnarray}
 F^{\mu \nu}_{~~~; \nu} + \frac{1}{2 \sqrt 3} \left( \sqrt{-g} \right)^{-1} 
   \epsilon ^{\mu \nu \rho \sigma \lambda } F_{\nu \rho } F_{\sigma \lambda } 
   = 0 \,. 
 \label{Maxeq}  
\end{eqnarray}
Then, as a solution, we obtain the following metric and the gauge potential 
one-form, respectively,   
\begin{eqnarray}
ds^2= -(dt+jr^2\sigma_3)^2 + dr^2 
      + \frac{r^2}{4}\left(\sigma_1^2+\sigma_2^2+\sigma_3^2\right) \,, 
\label{eq:Godel}
\end{eqnarray}
and 
\begin{eqnarray}
A=\frac{\sqrt{3}}{2}jr^2\sigma_3,\label{eq:Godel2} \,, 
\end{eqnarray}
where the one-forms $\sigma_i$ $(i=1,2,3)$ are given by 
\begin{eqnarray}
&&\sigma_1=\cos\psi d\theta+\sin\psi\sin\theta d\phi \,, \\
&&\sigma_2=-\sin\psi d\theta+\cos\psi\sin\theta d\phi \,, \\
&&\sigma_3=d\psi+\cos\theta d\phi \,, 
\label{def:inv-1-forms}
\end{eqnarray} 
with the coordinates $(r,\theta,\phi, \psi)$ having the ranges 
$0<r<\infty$, $0\le \theta<\pi$, $0\le \phi<2\pi$, $0\le \psi<4\pi$. 
The parameter, $j$, is called the {\sl G\"odel parameter}. 

\medskip 
The norm of the Killing vector $\psi^\mu := (\partial/\partial \psi)^\mu$ 
becomes negative in the region of $r>1/(2|j|)$ with the signature remaining 
Lorentzian, and therefore this solution admits closed timelike curves (CTCs), 
as the four-dimensional G\"odel universe does. 
The metric, eq.~(\ref{eq:Godel}), is completely homogeneous as a 
five-dimensional spacetime, just like the four-dimensional G\"odel universe 
is so in the four-dimensional sense. While the four-dimensional G\"odel 
universe is a spacetime filled with a pressureless perfect fluid balanced 
with a negative cosmological constant, 
the five-dimensional G\"odel universe given above, eqs.~(\ref{eq:Godel}) 
and (\ref{eq:Godel2}), is filled with a pressureless dust only, as 
immediately seen from its energy-momentum tensor 
\begin{eqnarray}
T^{\mu\nu}=\frac{1}{4\pi}\left(F^{\mu\nu}-\frac{1}{4}g^{\mu\nu}F^2\right)
          =\frac{3j^2}{\pi}(\partial/\partial t)^{\mu}(\partial/\partial t)^\nu \,,     
\end{eqnarray}
where it should be noted that the energy density itself is constant.

\subsection{A rotating GPS monopole}
The rotating GPS monopole solution is given by the following metric and 
the gauge potential one-form:  
\begin{eqnarray}
ds^2&=& -(dt+jr^2\sigma_3)^2 + k(r)^2dr^2 
        + \frac{r^2}{4}\left[k(r)(\sigma_1^2+\sigma_2^2)+\sigma_3^2\right] \,, 
\label{eq:rotaingGPS}
\\
A &=& \frac{\sqrt{3}}{2}jr^2\sigma_3 \,, 
\label{eq:rotaingGPS2}
\end{eqnarray}
where $k(r)$---called the {\sl squashing function}---is given by 
\begin{eqnarray}
k(r)=\frac{r_\infty^4}{(r_\infty^2-r^2)^2} \,.
\end{eqnarray}
It is immediate to see that the metric of the G\"odel universe, 
eq.~(\ref{eq:Godel}), is transformed as 
$dr\to k(r)dr$, $\sigma_1\to \sqrt{k(r)}\sigma_1$ and 
$\sigma_2 \to \sqrt{k(r)}\sigma_2$. 
By this transformation, the metric of the unit round $\rm S^3$ is 
deformed to the metric of a squashed $\rm S^3$ for which the radius 
of $\rm S^2$ is no longer equal to the radius of $\rm S^1$. 
(See the part of the metric in the square bracket of 
eq.~(\ref{eq:rotaingGPS}).) 
For this reason, this is called the squashing transformation. 
The range of the coordinate $r$ is $0<r<r_\infty$. 
The point $r=r_\infty$ turns out to correspond to spatial infinity.

\medskip 
Now let us introduce the following new radial coordinate 
\begin{eqnarray}
\rho = \frac{r_\infty r^2}{2(r_\infty^2-r^2)} \,, 
\label{def:rho:RotGPS}
\end{eqnarray}
so that $r\to r_\infty$ corresponds to $\rho \to \infty$. 
Then, the metric and the gauge potential one-form can be rewritten, 
respectively, as 
\begin{eqnarray}
 ds^2&=&-\left[ 
               dt + 4j\rho_0^2
               \left( 1+\frac{\rho_0}{\rho}\right)^{-1}\sigma_3 
         \right]^2 
\nonumber\\
  && + \left(1+\frac{\rho_0}{\rho}\right)
       \left[d\rho^2+\rho^2(\sigma_1^2+\sigma_2^2)\right] 
     +\rho_0^2\left(1+\frac{\rho_0}{\rho}\right)^{-1}\sigma_3^2 \,,
\label{eq:GPS}  
\end{eqnarray} 
\begin{eqnarray} 
  A=2\sqrt{3}\rho_0^2j\left(1+\frac{\rho_0}{\rho}\right)^{-1}\sigma_3 \,,\label{eq:GPSA}
\end{eqnarray}
where we have also introduced the constant $\rho_0=r_\infty/2$.

\medskip 
In this spacetime, depending on the choice of parameter range, 
we suffer from causality violation due to the existence of closed 
timelike curves (CTCs). 
(Recall that $\sigma_3$ in the first term of the above metric includes 
the periodic coordinate $\psi$.) 
To cure this, we hereafter restrict the range of the parameters 
$(j,\rho_0)$ as 
\begin{eqnarray}
j^2<\frac{1}{16\rho_0^2},\quad \rho_0>0 \,.
\label{rest:param:j-rho}
\end{eqnarray} 

\medskip 
When the G\"odel parameter is $j=0$, the metric, eq.~(\ref{eq:GPS}), 
coincides with the static GPS monopole solution 
given originally in Ref.~\cite{GP}. In this case it is immediate to see that 
the point of $\rho=0$ is a fixed point of the Killing vector field, 
$\psi^\mu =(\partial/\partial \psi)^\mu$, and the metric is analytic there, 
and thus the metric corresponds to a Kaluza-Klein monopole~\cite{GP}.  
A fixed point of some Killing field like this is often called a {\sl nut}. 
This is also the case even when $j \neq 0$.

\medskip 
Further, introduce new coordinates defined by 
\begin{eqnarray}
\bar t=\frac{t}{C},\quad \bar\psi=\psi-\frac{D}{C}\ t \,,
\end{eqnarray}
where the constants $C$ and $D$ are 
\begin{eqnarray}
  C=\sqrt{1-16j^2\rho_0^2} \,,\quad 
  D=\frac{4j}{\sqrt{1-16j^2\rho_0^2}} \,,  
\end{eqnarray} 
which make sense under the condition, eq.~(\ref{rest:param:j-rho}). 
For $\rho\to\infty$, the metric behaves as 
\begin{eqnarray}
ds^2 &\simeq& -d\bar t^2+d\rho^2+\rho^2(\sigma_1^2+\sigma_2^2)
              +\rho_0^2\left[1-16j^2\rho_0^2\right]\sigma_3^2 \,. 
\end{eqnarray} 
It is now clear that the metric is asymptotically locally flat 
and has the structure of a twisted $\rm S^1$ bundle over 
the four-dimensional Minkowski space-time. 
It is also clear that the presence of non-vanishing parameter $j$ means 
that although the spacetime, eq.~(\ref{eq:GPS}), has no black hole, it is 
{\sl rotating along the direction of the extra-dimension} 
specified by $\psi^\mu$.
We emphasize here again that CTCs which exist in the G\"odel universe, 
eq.~(\ref{eq:Godel}), now cease to exist as a result of the squashing 
transformation and the choice of the parameter range, 
eq.~(\ref{rest:param:j-rho}). 

\begin{figure}[!h]
\begin{center}
\includegraphics[width=0.5\linewidth]{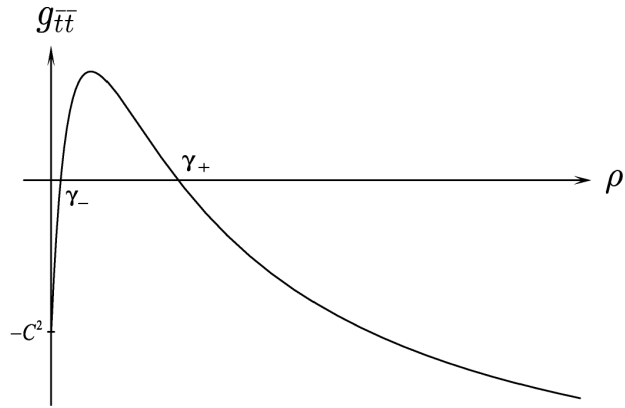}
\begin{minipage}{0.8\hsize}
\caption{The typical behavior of $g_{\bar t\bar t}$ in the case of 
$\sqrt{3}/8<|j|\rho_0<1/4$. There exists an ergoregion in the region 
such that $g_{\bar t \bar t}>0$. \label{fig:GPS_ergo}}
\end{minipage}
\end{center}
\end{figure}

\medskip 
Remarkably, this rotating GPS monopole solution possesses an ergoregion, 
despite the fact that there is no black hole event horizon in this spacetime. 
This can be seen as follows. 
The $\bar t\bar t$-component of the metric in the rest frame takes 
the following form near infinity,  
\begin{eqnarray}
 g_{\bar t\bar t} 
 = - \left[ 
          \left( 
                C+\frac{1-C^2}{C}\left(1+\frac{\rho_0}{\rho}\right)^{-1}
          \right)^2 
          -\frac{1-C^2}{C^2}\left(1+\frac{\rho_0}{\rho}\right)^{-1}
     \right] \,.
\end{eqnarray}
As shown in FIG.\ref{fig:GPS_ergo}, in the case of 
$\sqrt{3}/8<|j|\rho_0<1/4$, $g_{\bar t \bar t}$ becomes positive in 
the region of $\gamma_-<\rho<\gamma_+$, where 
\begin{eqnarray} 
\gamma_\pm:=\rho_0\frac{1-3C^2\pm(1-C^2)\sqrt{1-4C^2}}{2C^2} \,, 
\end{eqnarray}
and therefore there exists an ergoregion in that region, 
although there is no black hole horizon in the space-time. 
In a neighborhood of the nut, the ergoregion vanishes.

\medskip 
The energy-momentum tensor for the rotating GPS monopole solution, 
eqs.~(\ref{eq:rotaingGPS}) and (\ref{eq:rotaingGPS2}), is calculated as
\begin{eqnarray}
 T^{\mu\nu}=\frac{3 j^2}{\pi k(r)^2} 
            (\partial/\partial t)^\mu (\partial/\partial t)^\nu \,.  
\end{eqnarray}
Hence the rotating GPS monopole spacetime is filled with the pressureless 
perfect fluid with the same form of the energy-momentum tensor for 
the five-dimensional G\"odel universe. However, the energy-density 
is no longer constant, in contrast to the five-dimensional G\"odel universe, 
and falls off at large distances as the squashing function behaves 
$k(r_\infty) \rightarrow \infty \; (r \rightarrow r_\infty)$. 
Thus, the rotating GPS monopole solution asymptotically approaches 
a locally flat, vacuum spacetime.  
Furthermore, it can be shown by following Gauntlett et al.~\cite{Gauntlett} 
that the rotating GPS monopole is a supersymmetric solution 
on the self-dual Euclidean Taub-NUT base space (see Appendix).

\section{Black holes in a rotating GPS monopole}\label{sec:solution}
\subsection{Metric and gauge potential}
Now we 
present our new Kaluza-Klein black hole solutions 
to the equations (\ref{Eineq})-(\ref{Maxeq}) .
The metric and gauge potential are given, respectively, by 
\begin{eqnarray}
 ds^2 = - f(r)dt^2 - 2g(r)\sigma_3dt + h(r)\sigma_3^2 
      + \frac{k(r)^2dr^2}{V(r)} 
      + \frac{r^2}{4} 
        \left[k(r)\left(\sigma_1^2+\sigma_2^2 \right)+\sigma_3^2 \right] \,,
\label{def:metric:newKKBB}
\end{eqnarray}
and
\begin{eqnarray}
 A = \frac{\sqrt{3}}{2} 
     \left[\frac{q}{r^2}dt+\left(jr^2+2jq\right)\sigma_3\right] \,, 
\end{eqnarray}
where the metric functions, $f(r),g(r),h(r)$, and $V(r)$ are 
\begin{eqnarray}
&&f(r)=1-\frac{2m}{r^2}+\frac{q^2}{r^4} \,,\\
&&g(r)=jr^2+3jq \,,\\
&&h(r)=-j^2r^2(r^2+2m+6q) \,,\\
&&V(r)=1+\frac{-2m+16j^2(m+q)(m+2q)}{r^2}+\frac{q^2(1-8j^2(m+3q))}{r^4} \,,
\end{eqnarray}
and the squashing function $k(r)$ is 
\begin{eqnarray}
k(r)=\frac{V(r_\infty)r_\infty^4}{(r^2-r_\infty^2)^2} \,.
\end{eqnarray}
The one-forms $\sigma_i$ $(i=1,2,3)$ are given by eqs.~(\ref{def:inv-1-forms}).
The range of the ``spatial'' coordinates $(r,\psi,\phi,\theta)$ is 
$0<r<r_\infty$, $0\le\psi<4\pi$, $0\le\phi<2\pi$, and $0\le\theta<\pi$. 
The spacetime has the timelike Killing vector field, 
$\xi^\mu = (\partial /\partial t)^\mu$, and two Killing vector fields, 
$\psi^\mu$ and $\phi^\mu=(\partial /\partial \phi)^\mu$ with closed orbits.

\medskip 
The metric above can be viewed, on one hand, as a {\sl squashed} 
Reissner-Nordstr\"om-G\"odel black hole since the metric recovers that of 
the Reissner-Nordstr\"om-G\"odel black hole solution~\cite{Herdeiro} 
when one takes the limit $r_\infty\to\infty$ with keeping 
the other parameters finite, i.e., $k(r)\to 1$. 
(Note that $\psi^\mu$ can be timelike in this limit and CTCs appear.) 
On the other hand, if one takes the limit $m \rightarrow 0$ and 
$q \rightarrow 0$, then the metric recovers the rotating GPS monopole 
spacetime discussed in the previous sections and hence can be viewed 
as a charged black hole on the rotating GPS monopole background.

\begin{figure}[!h]
\begin{center}
\includegraphics[width=0.65\linewidth]{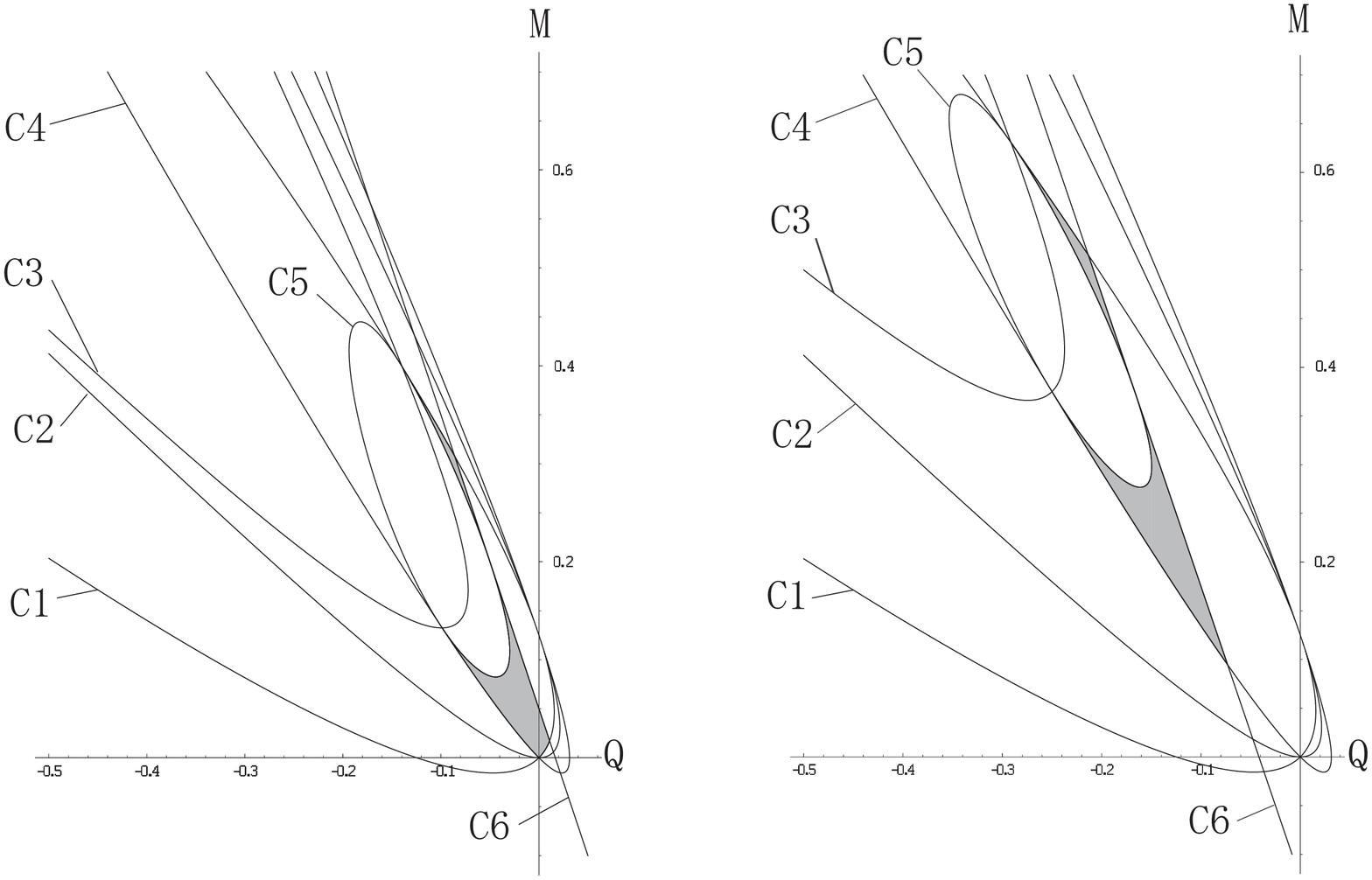}
\begin{minipage}{1.0\hsize}
\caption{The shaded regions in the left and right figures denote 
the parameter regions in the cases of $R_\infty^2=0.15$ and 
$R_\infty^2=0.50$, respectively.} 
\label{fig:parameter1}  
\end{minipage}
\end{center}
\vspace{1cm}
\begin{center}
\includegraphics[width=0.65\linewidth]{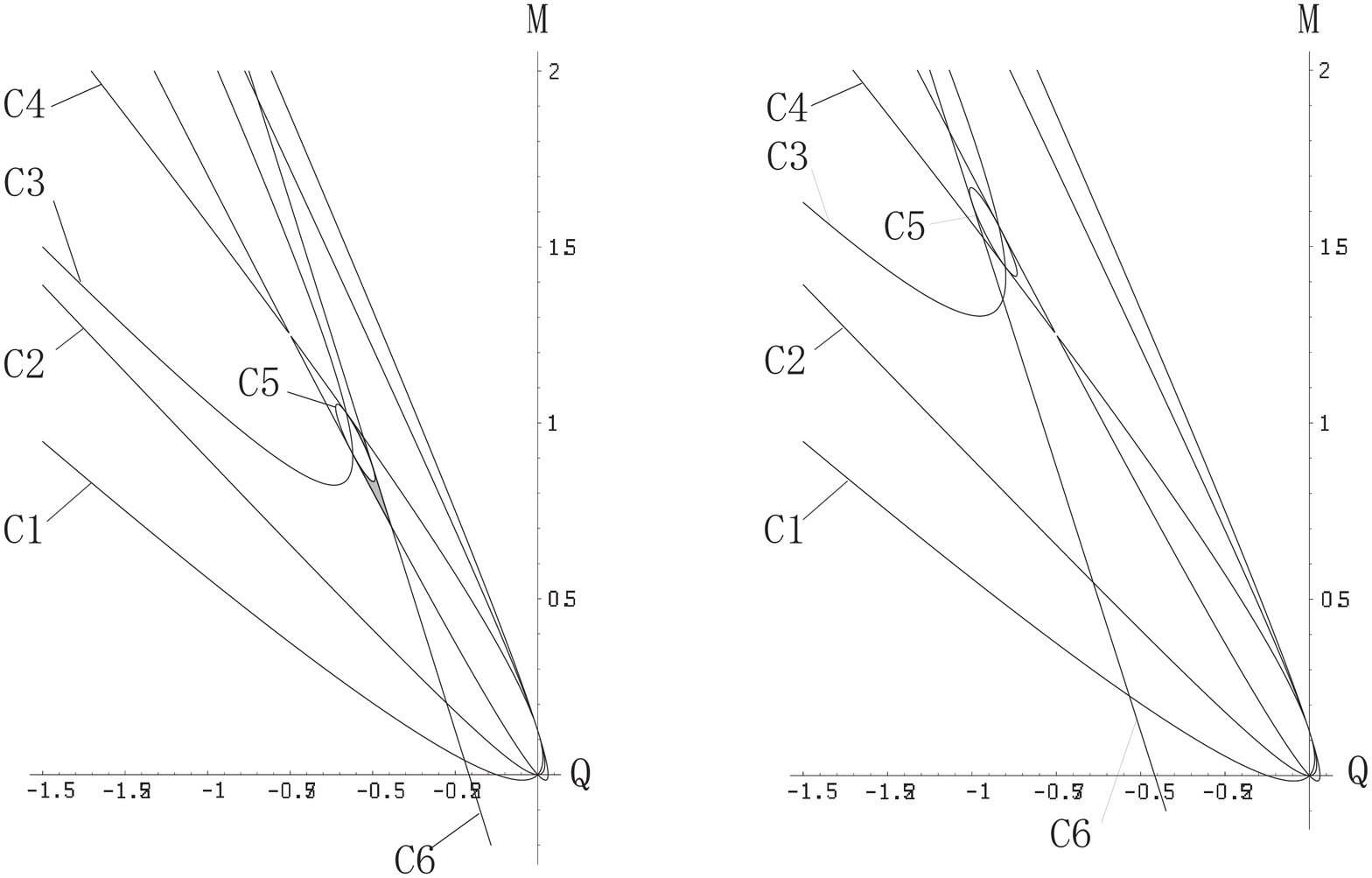}
\begin{minipage}{1.0\hsize}
\caption{The left and right figures correspond to the cases of 
$R_\infty^2=2.00$ and $R_\infty^2=3.00$, respectively. 
As $R_\infty$ increases, the curve $(C6)$, which comes from the condition 
for the absence of CTCs, shifts below. So if $R_\infty$ takes sufficiently 
large values, then CTCs appear outside the outer horizon and the 
parameter region disappears. } 
\label{fig:parameter2} 
\end{minipage}
\end{center}
\end{figure}

\medskip 
In what follows the 4 parameters $(j,m,q, r_\infty)$ that specify 
the above solutions are assumed to satisfy the following inequalities,   
\begin{eqnarray}
&&r_\infty^2>m-8j^2(m+q)(m+2q)>0 \,,\label{eq:para1} \\
&&1-8j^2(m+3q)>0\,,\label{eq:para2} \\
&&[m-q-8j^2(m+q)^2][m+q-8j^2(m+2q)^2]>0 \,,\label{eq:para3}\\
&&r_\infty^4+2[-m+8j^2(m+q)(m+2q)]r_\infty^2+q^2(1-8j^2(m+3q))>0 \,,
\label{eq:para4}\\
&&r_\infty^2<\frac{1}{4j^2}-(2m+6q) \,. 
\label{eq:para5}
\end{eqnarray} 
As we will see below, these inequalities are the necessary and 
sufficient conditions for the spacetime, eq.~(\ref{def:metric:newKKBB}), 
to admit two Killing horizons and no CTCs outside the horizon. 
The parameter region is shown in FIG.~\ref{fig:parameter1} 
and~\ref{fig:parameter2}, in which the parameters $(m,q,r_\infty)$ are  
normalized by $j$ as $M=j^2m,Q=j^2 q,R_\infty^2=j^2 r_\infty^2$, 
with the parameter $r_\infty$ being fixed. 
$C_i(i=1,\cdots,6)$ denote the following curves in 
the $(Q,M)$-planes:  
\begin{eqnarray} 
&&(C1)\ M-Q-8(M+Q)^2=0 \,,\\
&&(C2)\ M-8(M+Q)(M+2Q)=0 \,,\\
&&(C3)\ M-8(M+Q)(M+2Q)=R_\infty^2 \,,\\
&&(C4)\ M+Q-8(M+2Q)^2=0 \,,\\
&&(C5)\ R_\infty^4+2(-M+8(M+Q)(M+2Q))R_\infty^2+Q^2(1-8(M+3Q))=0 \,,\\
&&(C6)\ R_\infty^2=\frac{1}{4}-(2M+6Q) \,.
\end{eqnarray}

\subsection{Parameter region}  
Here we derive the parameter region (\ref{eq:para1})-(\ref{eq:para5}). 
The Killing horizons are located at the root $r>0$ of $V(r)=0$, i.e., 
the quadratic equations with respect to $r^2$: 
\begin{eqnarray}
r^4+[-2m+16j^2(m+q)(m+2q)]r^2+q^2(1-8j^2(m+3q))=0 \,. 
\label{eq:q2} 
\end{eqnarray}
This equation has two different positive real roots within the range 
of $(0,r_\infty^2)$ if and only if the parameters in the solution obey 
the inequalities 
\begin{eqnarray}
&&r_\infty^2>m-8j^2(m+q)(m+2q)>0 \,,\\ 
&&q^2(1-8j^2(m+3q))>0 \,,\\
&&r_\infty^4+[-2m+16j^2(m+q)(m+2q)]r_\infty^2+q^2(1-8j^2(m+3q))>0 \,,\\
&&[m-q-8j^2(m+q)^2][m+q-8j^2(m+2q)^2]>0 \,.
\end{eqnarray}
To avoid the existence of CTCs outside the horizons, we have to choose 
the parameters so that the two-dimensional part of the metric 
spanned by $(\psi,\phi)$ is positive-definite everywhere outside the outer 
horizon, i.e., $g_{\phi\phi}>0$ and $g_{\psi\psi}>0$. 
This is achieved whenever $g_{\psi\psi}>0$ 
since $g_{\phi\phi}=g_{\psi\psi}\cos^2\theta + (1/4)r^2k(r)\sin^2\theta$. 
For this, the following inequality needs to be satisfied everywhere 
in the region 
$(r_+^2,r_\infty^2)$,  
\begin{eqnarray}
h(r)+\frac{r^2}{4}>0\Longleftrightarrow u(r^2):=-4j^2r^2+(1-4j^2(2m+6q))>0 \,.
\end{eqnarray}
Since $u$ is a monotonically decreasing function of $r^2$, 
the necessary and sufficient condition for 
$u>0, \: r^2 \in (r_+^2,r_\infty^2)$ is simply $u(r_\infty^2)>0$. 
Hence the necessary and sufficient condition for the absence of CTCs 
outside the horizons is given by eq.~(\ref{eq:para5}).

\subsection{Asymptotic structure and asymptotic charges} 
In the coordinate system $(t,r,\theta,\phi,\psi)$, the limit 
$r \rightarrow r_\infty$ corresponds to spatial infinity. 
To investigate the asymptotic structure of the solution, 
as in the case of the rotating GPS monopole background, 
we introduce the new radial coordinate defined by 
\begin{eqnarray}
\rho=\rho_0\frac{r^2}{r_\infty^2-r^2} \,,
\end{eqnarray}
which is similar to eq.~(\ref{def:rho:RotGPS}) but the positive constant 
$\rho_0$ is now given by 
\begin{eqnarray}
\rho_0^2=\frac{r_\infty^2}{4} V(r_\infty) \,.
\end{eqnarray}
Moreover, we introduce coordinates $(\bar t, \bar \psi)$ so that the metric 
is in a rest frame at infinity
\begin{eqnarray}
dt=Ad \bar t \,,\quad d\psi=d\bar\psi+Bd \bar t \,,
\end{eqnarray}
where two constants $A$ and $B$ are chosen as
\begin{eqnarray}
A&=&\frac{\sqrt{1-4j^2(2m+6q+r_\infty^2))}r_\infty^2 
         }{ 
           \sqrt{(q^2 -2mr_\infty^2)(1-8j^2(m+3q))
           +r_\infty^4+32j^2q^2r_\infty^2 
                }
          } \,, 
\label{eq:A} \\ 
B&=&\frac{4j(3q+r_\infty^2)
         }{\sqrt{1-4j^2(2m+6q+r_\infty^2))
          }
   \sqrt{(q^2 -2mr_\infty^2)(1-8j^2(m+3q))+r_\infty^4+32j^2q^2r_\infty^2}}  
\,. 
\label{eq:B}
\end{eqnarray}
Then, for $\rho\to \infty$, the metric behaves as
\begin{eqnarray}
ds^2&\simeq&-d\bar t^2+d\rho^2+\rho^2(\sigma_1^2+\sigma_2^2)+L^2\sigma_3^2 \,,
\label{metric:asympt:L}
\end{eqnarray}
where the angular coordinate $\psi$ in $\sigma_3$ is replaced by $\bar\psi$ 
and $L$ is given by 
\begin{eqnarray}
 L^2=\frac{r_\infty^2}{4} \left[ 1-4j^2(2m+6q+r_\infty^2) \right] \,.  
\label{def:L} 
\end{eqnarray} 
The constant $L$ is {\sl real} under the condition, eq.~(\ref{eq:para5}), 
and describes the size of the extra dimension, as indicated from 
the asymptotic form of the metric, eq.~(\ref{metric:asympt:L}).  

\medskip 
The Komar mass and the Komar angular-momenta at spatial infinity are given, 
respectively, by 
\begin{eqnarray}
M_K &=& -3\pi
    \frac{ (mr_\infty^2-q^2)k^2-j^2 r_\infty^2(10q^2+8mr_\infty^2)k
           +128j^2q^2r_\infty^4+2j^2r_\infty^6 
         }{ 4\sqrt{
             ((q^2-2mr_\infty^2)k^2-2mr_\infty^2+r_\infty^4)(k-4j^2r_\infty^2)
                  } 
         } \,, 
\label{def:K-Mass}
\\ 
J_{\psi} &=& \frac{\pi}{4} 
        jr_\infty^2 
        \left[ 3q(1-8j^2(m+r_\infty^2)-72j^2q^2-4j^2r_\infty^4) \right] \,,
\label{J:psi} \\
J_{\phi} &=& 0 \,,
\label{J:phi}
\end{eqnarray}
where $k:=1-8j^2(m+3q)$. 
Therefore the spacetime has the 
angular momentum only in the direction of the extra-dimension, 
i.e., along $\psi^\mu$, as is the case of the rotating GPS monopole. 
It is clearly seen from eq.~(\ref{J:psi}) that the rotation with 
the angular momentum $J_\psi$---being proportional to $jr_\infty^2$---is 
a consequence of a combined effect of the background G\"odel rotation 
and the squashing transformation. 
The global charge is expressed as 
\begin{eqnarray}
 {\bf Q_e}&=&\frac{1}{4\pi}\int \left(*F-\frac{2}{\sqrt{3}}A\wedge F \right) 
  = -\frac{\pi\sqrt{3}}{2}q(1-8j^2(m+q)). \label{eq:Q}
\end{eqnarray}

\subsection{Near-Horizon Geometry and Regularity}  
As mentioned above, 
the hypersurfaces at $r=r_\pm$, where $V(r_\pm)=0$, are Killing horizons. 
To confirm this, we introduce new coordinates $(v,\psi')$ defined by 
\begin{eqnarray}
&&t=v+\int \frac{2\sqrt{h(r)+r^2/4}k(r)}{rV(r)}dr \,, 
\\
&&\psi=\psi'+ \int \frac{2k(r)g(r)}{rV(r)\sqrt{h(r)+r^2/4}}dr 
            + \frac{g(r_\pm)}{h(r_\pm)+r_\pm^2/4}v \,.
\end{eqnarray} 
In a neighborhood of $r=r_{\pm}$, the metric behaves as
\begin{eqnarray}
ds^2 \simeq  
     - \frac{ r_{\pm}k(r_{\pm}) 
            }{ 2\sqrt{h(r_\pm)+r_\pm^2/4}  
             }dvdr 
     + \left[ 
             \frac{r_{\pm}^2}{4}k(r_{\pm})(\sigma_1^2+\sigma_2^2) 
             + \left( h(r_\pm)+\frac{r_{\pm}^2}{4} \right) \sigma_3^2 
       \right] 
     + {\cal O}((r-r_{\pm})) \,, 
\end{eqnarray}
where the angular coordinate $\psi$ in $\sigma_3$ has been replaced with 
$\psi'$. The Killing vector field $v^\mu = (\partial/\partial v)^\mu $ 
becomes null on $r=r_\pm$ and, furthermore, it is straightforward to 
see that $v^\mu$ is hypersurface-orthogonal at $r=r_\pm$. 
Therefore the hypersurfaces $r=r_{\pm}$ are Killing horizons. 
Note also that in the coordinate system $(v,\phi,\psi',r,\theta)$, 
each component of the metric is analytic on and outside the black hole event 
horizon. Hence the space-time has no curvature singularity 
on and outside the black hole horizon.

\subsection{Shape of the horizons}
Now we investigate the shape of the horizons, especially, the ratio of 
$\rm S^2$ base space to $\rm S^1$ fiber. The horizon cross-section metric 
is 
\begin{eqnarray}
ds^2|_{r={\rm const.}} 
      = \frac{r^2}{4}k(r)(\sigma_1^2+\sigma_2^2) 
      + \left(h(r)+\frac{r^2}{4}\right)\sigma_3^2 \,. 
\end{eqnarray}
The squashing of $\rm S^3$ is denoted by the ratio $k(r)/(h(r)+r^2/4)$. 
In the case of $k(r_\pm)/(h(r_\pm)+r/4)>1$, the surface is called 
{\it oblate}, since the radius of ${\rm S}^2$ is larger than that of 
${\rm S}^1$. In the case of $k(r_{\pm})/(h(r_{\pm})+r_\pm^2/4)<1$, 
it is called {\it prolate}, since the radius of $\rm S^2$ is smaller than 
that of $\rm S^1$. 
In the case of $k(r_{+})/h(r_{+})=1$, it is called a round $\rm S^3$. 
In the static case~\cite{IM}, the outer horizon is always oblate. 
In contrast, in the rotating case~\cite{NIMT}, the horizon admits a prolate 
shape in addition to the round $\rm S^3$. In our solution, 
the outer horizon is always oblate, i.e., 
\begin{eqnarray}
\frac{r_+^2k(r_+)}{4h(r_+) + r_+^2} > 1 \,, 
\end{eqnarray}
and the inner horizon is always prolate, i.e., 
\begin{eqnarray}  
\frac{r_-^2k(r_-)}{4 h(r_-) + r_-^2} < 1 \,. 
\end{eqnarray}

\section{Special cases}\label{sec:case}  
\subsection{$j\to 0$} 
In the limit $j\to 0$, the solution coincides with the static charged 
Ishihara-Matsuno solution to the five-dimensional 
Einstein-Maxwell equations. In particular, in the case of 
$r_\infty\to \infty$, i.e., $k(r)\to 1$, the solution becomes 
the five-dimensional Reissner-Nordstr\"om solution, which is 
asymptotically flat in the standard five-dimensional sense.  
In the case $m=\pm q$, it has a degenerate horizon and is supersymmetric 
because it is included in a class of solutions on Taub-NUT base space 
in \cite{Gauntlett}, in which all purely bosonic supersymmetric solutions 
of minimal supergravity in five dimensions are classified. 
In fact, taking the limit of $m\to \pm q$ with introducing new coordinates 
$(\tilde t,\tilde r)$ and the parameters $(\tilde Q,R_{\infty})$ defined as 
\begin{eqnarray}
&&r^2=\frac{4(R_{\infty}^2 \tilde r+R_{\infty} \tilde Q)
           }{\tilde r+R_{\infty}} \,,\\
&&t=\frac{R_{\infty}^2}{R_{\infty}^2-\tilde Q}\tilde t \,,\\
&&r_\infty^2=4R_{\infty}^2 \,,\\
&&q=\pm4\tilde Q \,, 
\end{eqnarray}
we obtain the following metric 
\begin{eqnarray}
ds^2=-H^{-2}
                  d\tilde t^2+Hds_{\rm T-NUT}^2 \,. 
\label{def:metric:susybh}
\end{eqnarray}
Here $ds_{\rm T-NUT}^2$ is the metric on the Euclidean self-dual Taub-NUT 
space and is given, in terms of the Gibbons-Hawking coordinates, by 
\begin{eqnarray}
&&ds_{\rm T-NUT}^2=H_k(d\tilde r ^2+\tilde r ^2d\Omega_{S^2}^2)
                  +R_{\infty}^2H_k^{-1}\sigma_3^2 \,,  
\end{eqnarray}
where $H$ and $H_k$ are harmonic functions on the three-dimensional 
Euclid space and are expressed in the present coordinate system as 
\begin{eqnarray}
&&H=1+\frac{\tilde Q}{R_{\infty}\tilde r} \,,\\
&&H_k=1+\frac{R_{\infty}}{\tilde r} \,.
\end{eqnarray}
This metric, eq.~(\ref{def:metric:susybh}), coincides with the metric of 
the supersymmetric static black hole solutions with a compactified 
extra dimension on the Euclidean self-dual Taub-NUT space 
in Ref.~\cite{Gaiotto}. Moreover, in the case of 
$r_\infty\to \infty$, the metric can be written as 
\begin{eqnarray}
 ds^2&=&-\left(1\pm\frac{q}{R^2}\right)^{-2}dt^2 
       +\left(1\pm\frac{q}{R^2}\right)\left[dR^2+R^2d\Omega_{S^3}^2\right] \,,
\end{eqnarray}
where $R^2=r^2\mp q$ and $d\Omega_{S^3}^2$ is the metric on the unit round 
three-sphere. This is the metric of the extremal Reissner-Nordst\"om black 
hole solution which is supersymmetric solution on a four-dimensional Euclidean base space.

\subsection{Extremal case}
As shown previously, the Killing horizons are located at $r=r_\pm$.
In either the case $m-q-8j^2(m+q)^2=0$ or the case $m+q-8j^2(m+2q)=0$, 
the two horizons degenerate, i.e., $r_+=r_-$. Note however that 
as far as the G\"odel parameter $j$ is non-vanishing,  
the extremal limit of the present solutions are not supersymmetric 
in the sense of~\cite{Gauntlett}.

\subsection{Neutral cases}
For the Ishihara-Matsuno~\cite{IM} and the rotating Ishihara-Matsuno 
solution~\cite{NIMT}, the parameter, $q$, is simply in proportion to 
the physical charge $Q_e$. 
However, for the present solution, this is not the case 
due to the G\"odel parameter $j$, as seen in eq.~(\ref{eq:Q}).  
The physical global charge, $Q_e$, for the present solution 
vanishes in the case of either $q=0$ or $1-8j^2(m+q)=0$. 

\begin{figure}[!h]
  \begin{center}
   \includegraphics[width=0.35\linewidth]{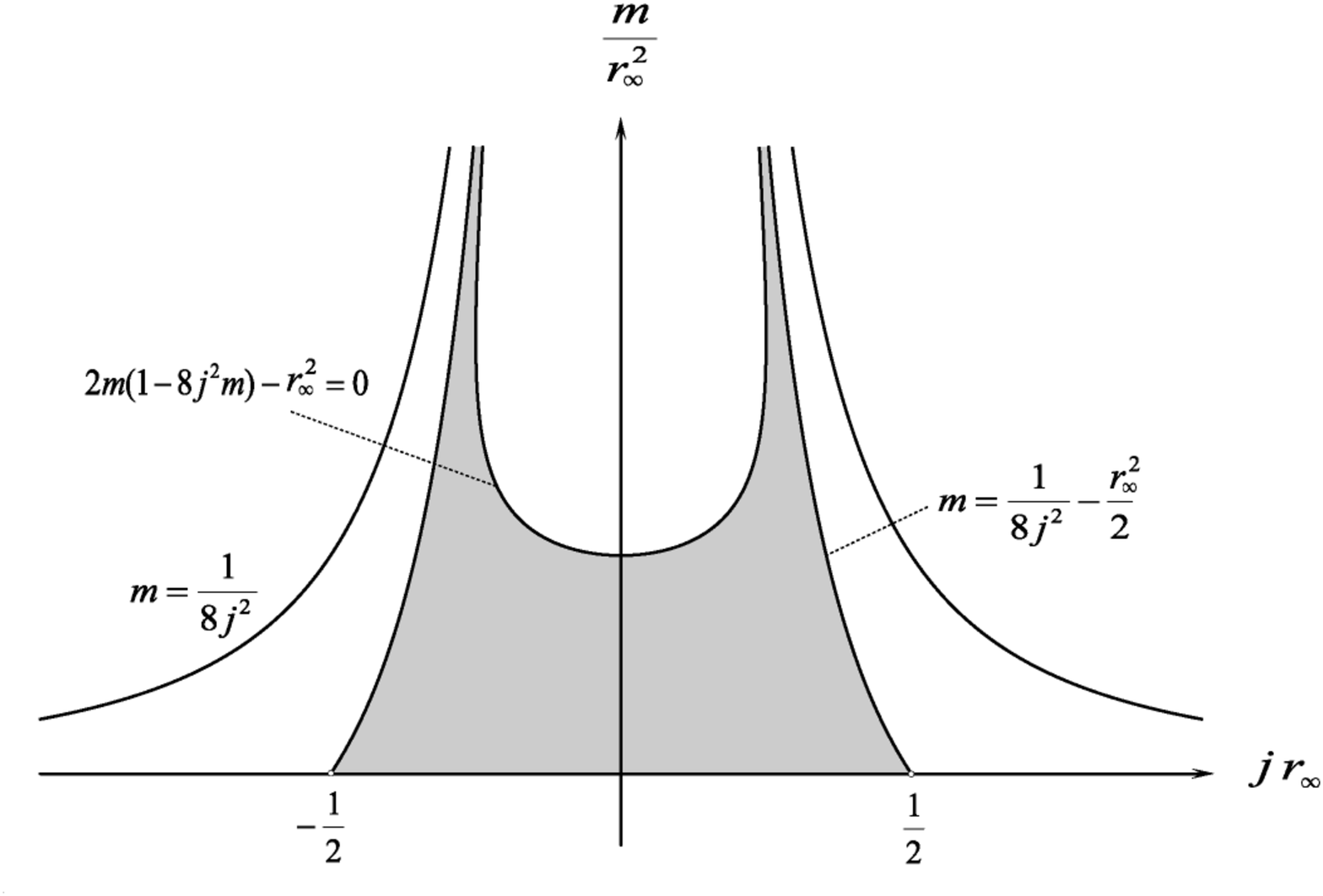}
 \begin{minipage}{0.7\hsize}
  \caption{Parameter region of the squashed Schwarzschild-G\"odel black hole solution in the $(j r_\infty,m/r_\infty^2)$-plane.}
  \label{fig:pararegion1}
 \end{minipage}
  \end{center}
 \vspace{0.7cm}
  \begin{center}
   \includegraphics[width=0.35\linewidth]{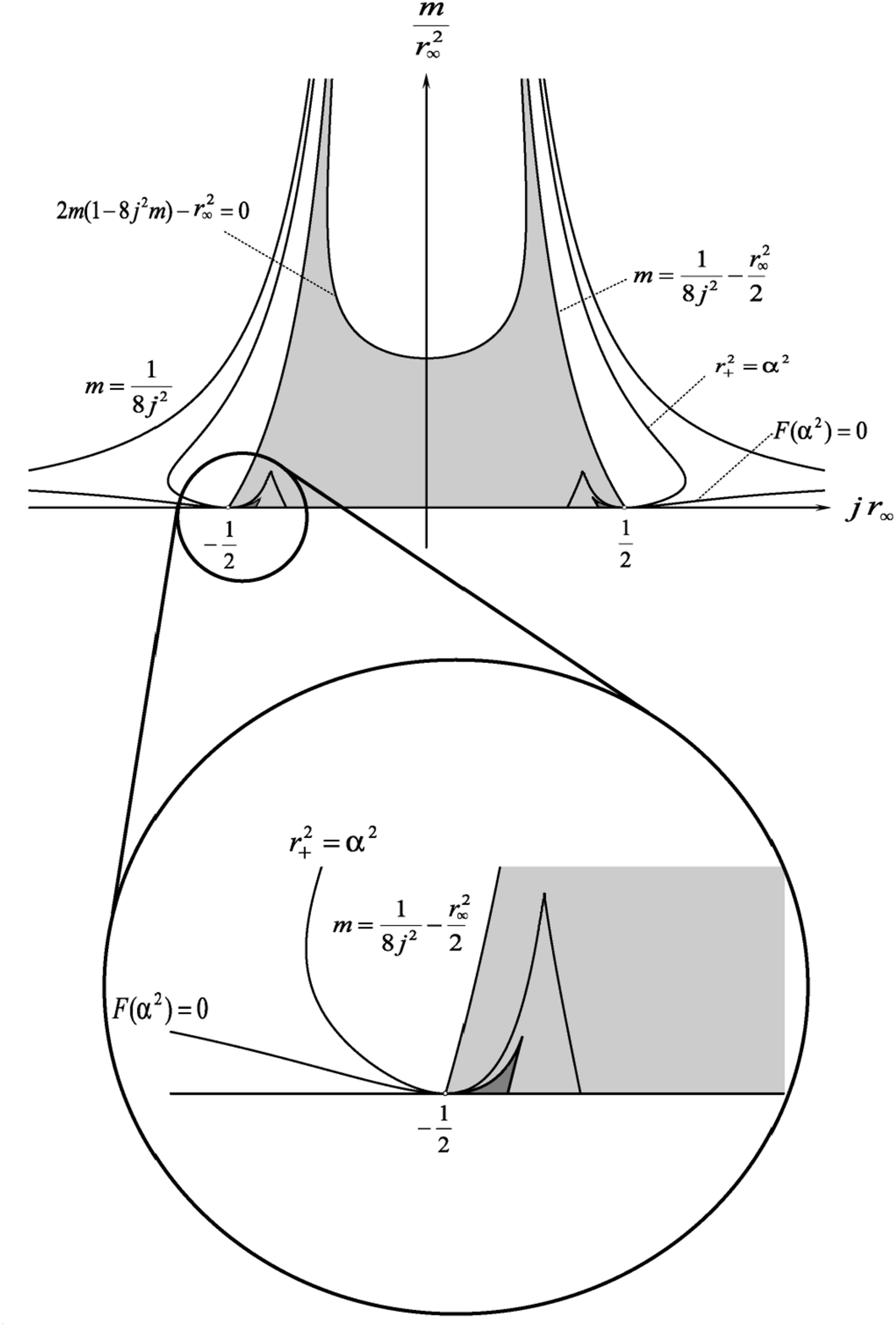}
  \end{center}
 \begin{minipage}{0.7\hsize}
  \caption{In the small dark parameter region in the above figure, there are two ergoregions. The below figure shows the close-up of the dark region in the above figure.}
  \label{fig:Sch_ergo}
 \end{minipage}
\end{figure}

\medskip 
For the former case ($q =0$), the solution coincides with the squashed 
Schwarzschild-G\"odel black hole solution~\cite{TIMN},   
which is obtained via the squashing transformation 
for the Schwarzschild-G\"odel black hole solution in 
Ref.~\cite{Gimon-Hashimoto}. 
In this case, the parameter region for the existence of a black hole horizon 
and non-existence of CTCs outside the horizon becomes 
\begin{eqnarray}
&& m>0 \,, \label{eq:pp1} \\
&& 1-8j^2m>0 \,,\label{eq:pp2}\\
&& -2m+\frac{1}{4j^2}>r_\infty^2>2m(1-8j^2m) \,. \label{eq:pp3} 
\end{eqnarray}
The horizon is located at $r= r_{\cal H}$ satisfying 
\begin{eqnarray}
r_{\cal H}^2=2m(1-8j^2m) \,.
\end{eqnarray}
The shaded region in FIG.\ref{fig:pararegion1} shows the region in which 
$(m,j)$ satisfy the inequalities (\ref{eq:pp1})-(\ref{eq:pp3}).
Ergoregions are located in the regions such that 
$F(r^2)=r^4g_{\bar t \bar t}$ is positive, where the function $F(r^2)$ is 
a cubic equation with respect to $r^2$, and is explicitly written as 
\begin{eqnarray}
F(r^2)&=&-16j^4r^8_\infty r^6-4j^2r_\infty^8(1-8j^2(m+r^2_\infty))r^4 
\nonumber \\
&& - \left[ 
           r_\infty^4(1-4j^2(2m+r^2_\infty)) 
     \right]^2r^2 
   + 2mr_\infty^8 \left[ 1-2j(2j(2m+r^2_\infty)) \right]^2 \,.
\end{eqnarray}
Note that $g_{\bar t\bar t}(r=r_{\cal H})>0$ and 
$g_{\bar t\bar t}(r=r_\infty)<0$ always hold. Hence the boundary of 
the ergoregion is always located at $r$ such that $g_{\bar t\bar t}=0$. 
Interestingly, when $F(\alpha^2)<0$ and $r_{\cal H}<\alpha$, 
there exist two ergoregions $r_{\cal H}<r<r_1$ and $r_2<r<r_3$ 
outside the black hole horizon, where $\alpha^2$ and $\beta^2$ 
($0<\alpha<\beta$) are the roots of the quadratic equation with respect to 
$r^2$,  $\partial_{r^2}F(r^2)=0$, and $r_i^2\ (i=1,2,3, \ r_1<r_2<r_3)$ 
are the roots of the cubic equation with respect to $r^2$, $F(r^2)=0$. 
The small dark region in FIG.\ref{fig:Sch_ergo} denotes the set of the 
solutions which admit two ergoregions outside the horizon.

\begin{figure}[!h]
\begin{center}
\includegraphics[width=0.8\linewidth]{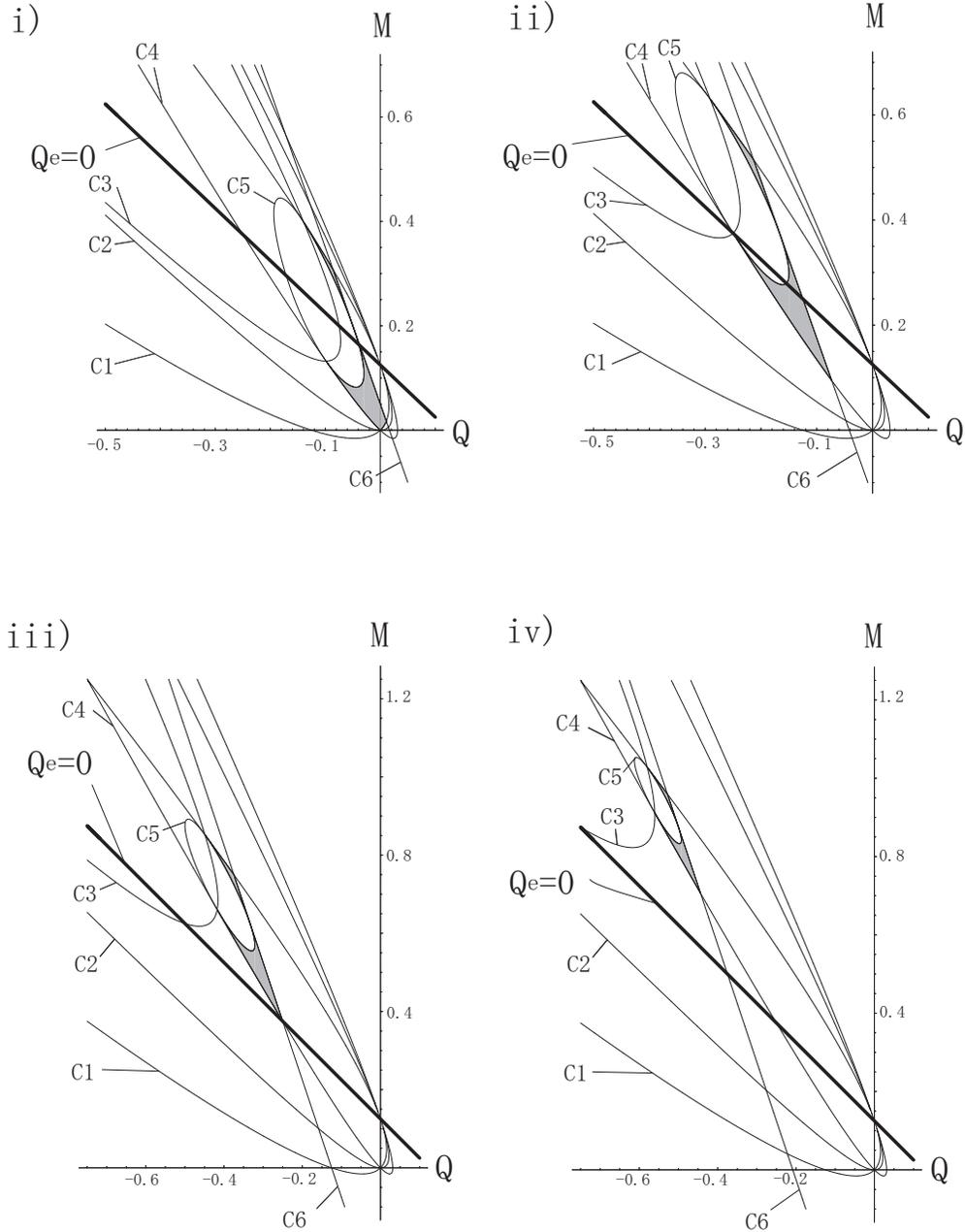}
\begin{minipage}{1.0\hsize}
\caption{The bold lines show the lines where the global charge 
$Q_e$ vanishes for $(i)\ R_\infty^2=0.15, \: 
(ii)\ R_\infty^2=0.50, \: 
(iii)\ R_\infty^2=1.00, \:  
(iv)\ R_\infty^2=1.50$. 
\label{fig:Q}}
\end{minipage}
\end{center}
\end{figure}

\medskip 
For the latter case (i.e., $1-8j^2(m+q)=0$), 
FIG.~\ref{fig:Q} shows the line of $1-8(M+Q)=0$ 
in the $(Q,M)$-plane for the four values of $R_\infty^2=0.15,0.50,1.00$, $1.50$. 
In the case $R_\infty^2=0.15,0.50,1.00$, the line, $1-8(M+Q)=0$, 
penetrates through the allowed parameter region, whereas in the case 
$R_\infty^2=1.50$, it does not.

\bigskip \bigskip

\vspace{1cm}

\section{Summary and Conclusions}\label{sec:summary}

In this paper, we have presented a new class of stationary charged 
Kaluza-Klein black hole solutions, which are constructed from 
the Reissner-Nordst\"om-G\"odel black hole solution 
in five-dimensional Einstein-Maxwell-Chern-Simons theories, 
via the squashing transformation.  
Our solutions can be viewed as a charged black hole in the rotating 
GPS monopole background and be seen to be rotating by the effect of 
the rotation of the background GPS monopole.  
(Note that the rotation of the GPS background itself is caused by 
the existence of the non-homogeneous electromagnetic field.)  
Like known charged black hole solutions, our black hole spacetime 
has two horizons, the outer and inner horizons. 
Although the cross-section geometry of the outer horizon is of squashed 
$\rm S^3$ (hence essentially a five-dimensional black hole), 
at large distances the spacetime behaves effectively as a four-dimensional 
spacetime (provided that the scale of extra-dimension, 
$L$ given by eq.~(\ref{def:L}), is sufficiently small compared 
to other physically relevant scales at large distances).  

\medskip 
Some further remarks are in order. 
Our solutions are specified by four parameters, $(m,q,r_\infty,j)$.  
When the G\"odel parameter, $j$, vanishes, the solution becomes the 
Ishihara-Matsuno black hole solution, which is a static charged Kaluza-Klein 
black hole solution in the five-dimensional Einstein-Maxwell theory, 
in which the parameters, $m$ and $q$, are directly related to the globally 
defined mass and the electric charge, respectively, 
and $r_\infty$ is proportional to the size of the extra-dimension.   
When both the $m$ and $q$ are zero, our solution 
reduces to the rotating GPS monopole solution.

\medskip 
It should be noted that 
the parameter $r_\infty$ has an upper bound which comes from the condition 
for the absence of closed timelike curves. 
This is in contrast to the fact that 
neither the squashed Kerr-G\"odel black hole solution~\cite{TIMN} 
nor the rotating Ishihara-Matsuno solution~\cite{NIMT} 
has CTCs (outside the black hole horizon) for any $r_\infty < \infty$. 
\footnote{  
Note that the {\sl squashed Kerr-Newman G\"odel black hole 
solution}---which has 5 parameters $m,q,j,r_\infty$ and the angular-momentum 
parameter $a$---has recently been constructed 
in~\cite{TIMN}. 
However, it would not appear to be straightforward to specify possible 
parameter region for which CTCs cease to exist outside the black hole 
horizon of this solution.   
} 

\medskip 
In the Ishihara-Matsuno solution~\cite{IM}, the $m=\pm q$ case  
corresponds to a supersymmetric solution, i.e., the static black hole 
on Taub-NUT space~\cite{Gauntlett}. 
However, in the rotating Ishihara-Matsuno solution~\cite{NIMT}, 
the two cases, $m=-q$ and $m=q$, describe different solutions due to 
the existence of a Chern-Simons term. 
Only the case of $m=-q$ corresponds to a supersymmetric Kaluza-Klein black 
hole solution of Gaiotto {\it et al}~\cite{Gaiotto}. In our solution, 
neither the case $m=-q$ nor the case $m=q$, is supersymmetric and 
the two horizons do not degenerate. For either $m-q=8j^2(m+q)^2$ 
or $m+q=8j^2(m+2q)$, the outer and inner horizons degenerate, but 
in either case our solution is non-supersymmetric.

\medskip 
Finally we would like to make two comments on the squashing 
transformation. 
First, we recall that our starting-point geometry (either the 
Reissner-Nordst\"om-G\"odel black hole solution or the five-dimensional 
G\"odel universe) suffers from causality violation, but nevertheless 
our resulting solutions do not. In view of this, we can anticipate that 
the squashing transformation may be used to generate much larger variety of 
physically interesting solutions, starting not only from physically sensible 
geometries (e.g., those of asymptotically flat, with no CTCs, etc) but 
even from various causally or metrically {\sl pathological} geometries 
such as spacetimes with CTCs and/or naked singularities. 
Second, as far as the present authors know, the squashing transformation 
has so far been applied only to cohomogeneity-one black hole solutions 
such as five-dimensional Myers-Perry black hole solutions with two equal 
angular momenta and the Cvetic-Lu-Pop's charged black hole solutions  
with two equal angular momenta. 
It would be interesting to consider a generalization of the squashing 
technique so that it would apply to cohomogeneity-two spacetimes 
such as black ring~\cite{ER}, black holes with two 
rotations~\cite{MP}, black lens~\cite{LMP,LMP2}, or even wider class 
of spacetimes.

\bigskip 
\begin{center}
{\bf Acknowledgements} 
\end{center}  
AI wishes to thank the Perimeter Institute for Theoretical Physics 
for its hospitality during the time some of this research was carried out. 
This research was supported in part by the Japan Society for the Promotion 
of Science and in part by Perimeter Institute for Theoretical Physics.

\appendix

\section{Supersymmetric rotating GPS monopole}
Here we show that our background rotating GPS monopole solution is 
supersymmetric, using the results of Ref.~\cite{Gauntlett}. 
We first recall that according to \cite{Gauntlett}, all supersymmetric 
solutions of the five-dimensional minimal supergravity have a non-spacelike 
Killing vector field, and 
when the Killing vector field $\partial/\partial t$ is timelike,
the metric and the gauge potential are given, respectively, by 
\begin{eqnarray}
ds^2=-H^{-2}(dt+\bm\omega)^2+Hds^2_{{\cal B}} \,, 
\ {\bm A}=\frac{\sqrt{3}}{2}[H^{-1}(dt+\bm\omega)-\bm\beta] \,, 
\label{eq:SUSY}
\end{eqnarray}
where $ds^2_{\cal B}$ is a metric of a hyper-K\"ahler space ${\cal B}$. 
The scalar function $H$, one-forms $\bm\omega$ and $\bm\beta$ on ${\cal B}$ 
are given by 
\begin{eqnarray}
\Delta H=\frac{4}{9}(G^+)^2 \,,\quad 
dG^+=0 \,,\quad 
d{\bm \beta}=\frac{2}{3}G^+ \,. 
\label{eq:SUSY2} 
\end{eqnarray}
Here, $\triangle$ is the Laplacian on ${\cal B}$ and the two-form $G^+$ is 
the self-dual part of the one-form $H^{-1}{\bm \omega}$, given by 
\begin{eqnarray}
  G^+:=\frac{1}{2}H^{-1}(d\omega+*d\omega) \,,  
\label{eq:SUSY3}  
\end{eqnarray}
where $(G^+)^2:=\frac{1}{2}G_{mn}G^{mn}$ and $*$ is the Hodge dual operator on 
${\cal B}$. Since $\partial/\partial t$ is a Killing vector field associated 
with time translation, all components are independent of 
the time coordinate $t$. 

\medskip  
We now show that the rotating GPS monopole solution takes the form 
(\ref{eq:SUSY})-(\ref{eq:SUSY3}). 
Comparing eq.~(\ref{eq:GPS}) and eq.~(\ref{eq:SUSY}), we can read off 
the correspondence of these two metrics as follows 
\begin{eqnarray}
H=1 \,,\quad {\bm \omega}=4jV\sigma_3 \,,
\end{eqnarray}
\begin{eqnarray}
ds^2_{\cal B}=V^{-1}(dr^2+r^2d\Omega_{S^2}^2)
             +V \rho_0^2(d\psi+\cos\theta d\phi)^2 \,,
\end{eqnarray}
with 
\begin{eqnarray}
V^{-1}=1+\frac{\rho_0}{\rho} \,,
\end{eqnarray}
where $ds^2_{\cal B}$ is the metric of the Euclidean self-dual Taub-NUT space 
and $V^{-1}$ is a harmonic function on the three-dimensional Euclid 
space ${\mathbb E}^3$. 
Note that $d\sigma_3=-\sigma_1\wedge \sigma_2$. 
So the exterior derivative of the one-form $\bm \omega$ can be written as 
\begin{eqnarray}
d\bm \omega &=& 4j\left(\frac{\rho_0}{\rho^2}V^2d\rho\wedge\sigma_3
                        -V\sigma_1\wedge \sigma_2 
                  \right) \,.
\end{eqnarray}
Then the dual of the one-form $\rm \omega$ is obtained as
\begin{eqnarray}
*d\bm \omega&=&-4j\left(
                        \frac{\rho_0}{\rho^2}V^2d\rho\wedge\sigma_3 
                        -V\sigma_1\wedge \sigma_2 
                  \right) \,.
\end{eqnarray}
As a result, we find 
\begin{eqnarray}
G^+=0 \,.
\end{eqnarray}
The second equation eq.~(\ref{eq:SUSY2}) is automatically satisfied. 
The first equation of eq.~(\ref{eq:SUSY2}) becomes 
\begin{eqnarray}
\Delta_{T-NUT}H=0 \,,   
\end{eqnarray}
and obviously, $H=1$ solves this equation. 
The third equation of eq.~(\ref{eq:SUSY2}) yields $d\bm\beta=0$. 
Hence the one-form $\bm \beta$ takes a closed form at least locally, 
i.e., there exists some function $\gamma $ such that $\bm \beta=d\gamma$. 
We can set the function $\gamma$ to zero by the gauge transformation. 
Thus the rotating GPS monopole solution (\ref{eq:GPS})-(\ref{eq:GPSA}) 
is supersymmetric.

\end{document}